%
%
\def\today{\ifcase\month\or January\or February\or March\or April\or May\or
June\or July\or August\or September\or October\or November\or December\fi
\space\number\day, \number\year}
%
%
\newcount\notenumber

\def\note{\global\advance\notenumber by 1 \footnote{$^{\the\notenumber}$}}
%
%
\newif\ifsectionnumbering
\newcount\eqnumber
\def\cleareqnumber{\eqnumber=0}
\def\numbereq{\global\advance\eqnumber by 1
\ifsectionnumbering\eqno(\the\secnumber.\the\eqnumber)\else\eqno
(\the\eqnumber)\fi}
\def\eqalinno{{\global\advance\eqnumber by 1}
\ifsectionnumbering(\the\secnumber.\the\eqnumber)\else(\the\eqnumber)\fi}
\def\name#1{\ifsectionnumbering\xdef#1{\the\secnumber.\the\eqnumber}
\else\xdef#1{\the\eqnumber}\fi}
\def\nosectionnumbering{\sectionnumberingfalse}
\sectionnumberingtrue
%
%
\newcount\refnumber

\immediate\openout1=refs.tex
\immediate\write1{\noexpand\frenchspacing}
\immediate\write1{\parskip=0pt}
\def\ref#1#2{\global\advance\refnumber by 1%
[\the\refnumber]\xdef#1{\the\refnumber}%
\immediate\write1{\noexpand\item{[#1]}#2}}
\def\tie{\noexpand~}

%
%
\font\twelvebf=cmbx10 scaled \magstep1
\newcount\secnumber

\def\newsection#1.{\ifsectionnumbering\cleareqnumber\else\fi%
        \global\advance\secnumber by 1%
        \bigbreak\bigskip\par%
        \line{\twelvebf\the\secnumber. #1.\hfil}\nobreak\medskip\par\noindent}
%
%
%
\def \sqr#1#2{{\vcenter{\vbox{\hrule height.#2pt
        \hbox{\vrule width.#2pt height#1pt \kern#1pt
                \vrule width.#2pt}
                \hrule height.#2pt}}}}
\def\Box{{\mathchoice\sqr54\sqr54\sqr33\sqr23}\,}
%
%
%
\newdimen\fullhsize
\def\fiddle{\fullhsize=6.5truein \hsize=3.2truein}
\def\fullline{\hbox to\fullhsize}
\def\mkhdline{\vbox to 0pt{\vskip-22.5pt
        \fullline{\vbox to8.5pt{}\the\headline}\vss}\nointerlineskip}
\def\mkftline{\baselineskip=24pt\fullline{\the\footline}}
\let\lr=L \newbox\leftcolumn
\def\twocolumns{\fiddle
        \output={\if L\lr \global\setbox\leftcolumn=\columnbox
                \global\let\lr=R \else \doubleformat \global\let\lr=L\fi
                \ifnum\outputpenalty>-20000 \else\dosupereject\fi}}
\def\doubleformat{\shipout\vbox{\mkhdline
                \fullline{\box\leftcolumn\hfil\columnbox}
                \mkftline} \advancepageno}
\def\columnbox{\leftline{\pagebody}}
\nosectionnumbering
\magnification=1200
\def\pr#1 {Phys. Rev. {\bf D#1\tie }}
\def\pe#1 {Phys. Rev. {\bf #1\tie}}
\def\pre#1 {Phys. Rep. {\bf #1\tie}}
\def\pl#1 {Phys. Lett. {\bf B#1\tie }}
\def\prl#1 {Phys. Rev. Lett.. {\bf #1\tie }}
\def\np#1 {Nucl. Phys. {\bf B#1\tie }}
\def\ap#1 {Ann. Phys. (NY) {\bf #1\tie }}
\def\cmp#1 {Commun. Math. Phys. {\bf #1\tie }}
\def\imp#1 {Int. Jour. Mod. Phys. {\bf A#1\tie }}
\def\mpl#1 {Mod. Phys. Lett. {\bf A#1\tie}}
\def\tie{\noexpand~}

\parskip=15pt plus 4pt minus 3pt
\headline{\ifnum \pageno>1\it\hfil Massive Higher Spin
States $\ldots$\else \hfil\fi}
\font\title=cmbx10 scaled\magstep1
\font\tit=cmti10 scaled\magstep1
\footline{\ifnum \pageno>1 \hfil \folio \hfil \else
\hfil\fi}
\raggedbottom


\def\half{{\textstyle{1\over2}}}
\def\ha{{1\over2}}
\overfullrule0pt


\rightline{\vbox{\hbox{CERN-TH/98-304}\hbox{NYU-TH/98/08/01}\hbox{RU98-10-B }
\hbox{hep-th/9809142}}}
\vfill
\centerline{\title MASSIVE HIGHER SPIN STATES IN STRING THEORY AND}
\centerline{\title THE PRINCIPLE OF EQUIVALENCE}
\vfill
{\centerline{\title Ioannis Giannakis${}^{a,b}$,
James T.~Liu${}^{a}$ and Massimo Porrati${}^{a,b,c}$ \footnote{$^{\dag}$}
{\rm e-mail: \vtop{\baselineskip12pt
\hbox{giannak@theory.rockefeller.edu, jtliu@theory.rockefeller.edu,}
\hbox{massimo.porrati@nyu.edu}}}}
}
\medskip
\centerline{$^{(a)}${\tit Physics Department, The Rockefeller
University}}
\centerline{\tit 1230 York Avenue, New York, NY
10021-6399}
\medskip
\centerline{$^{(b)}${\tit Department of Physics, New York University}
\footnote{$^*$}{\rm Permanent address.}}
\centerline{\tit 4 Washington Pl., New York, NY 10003}
\medskip
\centerline{$^{(c)}${\tit Theory Division, CERN}}
\centerline{\tit CH 1211 Geneva 23 (Switzerland)}
\vfill
\centerline{\title Abstract}
\bigskip
{\narrower\narrower\noindent
In this paper we study three point functions of the Type II
superstring involving one graviton and two massive states, focusing in
particular on the spin-7/2 fermions at the first mass level.  Defining a
gravitational quadrupole ``$h$-factor'', we find that the non-minimal
interactions of string states in general are parametrized by $h\ne1$, in
contrast to the preferred field theory value of $h=1$ (for tree-level
unitarity).  This difference arises from the fact that consistent
gravitational interactions of strings are related to the presence of
a complete tower of massive states, not present in the ordinary
field theory case.
\par}
\vfill\vfill
\vbox{\hbox{CERN-TH/98-304}\hbox{September 1998}}
\break


\newsection Introduction.%
While the matter of consistent interactions of massive higher spin fields
with gravity has been fairly well studied in the context of field theory
\ref\deser{C. Aragone and S. Deser, \pl86 (1979) 161; M. Duff and S. M.
Christensen, \np154 (1979) 301.},
less is known in the case of string theory.  Although for the latter, one
may argue that strings must {\it a priori} present a consistent theory of
gravity, it is nevertheless instructive to examine the nature of such
interactions and to determine in particular how strings achieve such a
consistency.

The simplest recipe for coupling massive fields to gravity (the ``minimal
coupling'') is inherently ambiguous. Indeed, by replacing ordinary
derivatives with covariant derivatives one is still free to add to the
action terms that vanish on flat space; the commutator of two covariant
derivatives, for instance.
In Ref.~%
\ref\porrati{M. Porrati, \pl304 (1993) 77; A. Cucchieri, M. Porrati and
S. Deser, \pr51 (1995) 4543.}
a prescription was given for fixing some of the ambiguities. There,
by imposing tree-level unitarity up to the Planck scale on forward
``Compton'' scattering amplitudes of a single massive high-spin
field, many coefficients in its action were fixed unambiguously.
In particular, it was shown that starting from spin-3/2, tree-level
unitarity requires the presence of terms proportional to the Riemann
tensor in the three-point vertex describing single-graviton emission
and absorption by the fermions.

Tree level unitarity is
really the statement that an interacting theory is weakly coupled up to a
certain energy scale, $M_{Planck}$ or $(\alpha')^{-1/2}$, for instance.
The recipe of Ref.~[\porrati]
is not immediately applicable to string theory.
The spectrum of string theory, indeed, contains many particles of
ever-increasing spin, some of which degenerate in mass. Thus, for instance,
tree-level unitarity of a massive high-spin particle is achieved in string
theory in part because of the three-point vertex, but also because an infinite
number of particles of ever-increasing mass and spin propagate as
intermediate states. To find which one-graviton vertex is selected by string
theory, one must therefore resort to direct calculation.

The computation of the three-point vertex involving one graviton and two
spin-7/2 fermions in superstring theory is described in Section 2, which
also exhibits the effective action reproducing the string amplitudes.
Section 3 focuses on the ``gravitational quadrupole'' term, that is the
three-point vertex proportional to the Riemann tensor.
In Section 3 we compare the results obtained from superstring theory with the
predictions
of tree-level unitarity. Section 4 extends the computation of the three-point
vertex and effective action to states of arbitrary spin, and contains
some comments on the implications of our results for tree-level unitarity in
strings.

\newsection Massive three-point functions.%
We begin by examining the simplest massive string interactions of the Type
II string, namely three point functions at the first mass level.  In this
case, the (ten-dimensional) closed superstring states fall into the
$SO(9)$ spin representations%
\note{Both IIA and IIB theories have identical massive spectra;
their interactions, however, are in general distinct.}
$[44+84+128]_L\times[44+84+128]_R$.  Correspondingly, in four dimensions,
such states carry spins up to four.  Focusing on spacetime fermions ({\it
i.e.}~the R-NS sector), we now describe the vertex operator for emission
or absorption of states in the $128\times44$ of $SO(9)$ and in the
$(q,{\overline q})=(-{1\over 2},-1)$ ghost picture.

Massive vertex operators have been discussed previously in
\ref\pro{I. G. Koh, W. Troost and A. Van Proeyen, \np292 (1987) 201;
Y. Tanii and Y. Watabiki, \np316 (1989) 171;
S. Weinberg, \pl156 (1985) 309.}.
Working in the R-NS sector, the vertex operator
may be written as
$$
V_{F}^{(-{1\over 2}, -1)}(z, {\overline z})=u^{\alpha}_{\mu\nu\rho}(X)
S_{\alpha}e^{-{{\phi}\over 2}}{\partial}X^{\mu}e^{-{\overline{\phi}}}
{\overline{\psi}}^{\nu}
{\overline{\partial}}X^{\rho}+{\upsilon}^{{\dot{\alpha}}}_{\mu
\nu\rho}(X)({\gamma}_{\lambda})_{{\dot{\alpha}}}^{\beta}{\psi}^{\mu}
{\psi}^{\lambda}S_{\beta}e^{-{{\phi}\over 2}}
e^{-{\overline{\phi}}}{\overline{\psi}}^{\nu}
{\overline{\partial}}X^{\rho}.
\numbereq\name{\eqena}
$$
In this expression $S_{\alpha}$ and $e^{-{{\phi}\over 2}}$ represent
the spin fields for the two-dimensional fermions ${\psi}^{\mu}$ and
the superconformal ghosts $\beta, \gamma$ respectively.
The wavefunctions $u$ and $v$ are now constrained by demanding that
$V_{F}^{(-{1\over 2}, -1)}(z, {\overline z})$ is BRST invariant,
{\it i.e.}~that $[ Q, V_{F}^{(-{1\over 2}, -1)}(z, {\overline z})]$
and similarly $[ {\overline Q}, V_{F}^{(-{1\over 2}, -1)}(z, {\overline z})]$
vanish up to a total derivative. The BRST charge for the Type II
string theory is given by
$$
\eqalign{
Q&={\int}dzc(z)(-{1\over 2}{\eta^{\mu\nu}}{\partial}X_{\mu}
{\partial}X_{\nu}-{1\over 2}{\eta^{\mu\nu}}{\psi_{\mu}}{\partial}{\psi_{\nu}}
-{3\over 2}{\beta}{\partial}{\gamma}-{1\over 2}
{\partial}{\beta}{\gamma})(z) \cr
&+{\int}d{z}(bc{\partial}c
+{1\over 2}{\gamma}{\eta}^{\mu\nu}{\psi_{\mu}}{\partial}X_{\nu}
-{1\over 4}b{\gamma}^{2})(z), \cr}
\numbereq\name{\eqasoud}
$$
and a similar expression for $\overline Q$.
We find that $[ Q, V_{F}^{(-{1\over 2}, -1)}(z, {\overline z})]=
{\partial}(c(z)V_{F}^{(-{1\over 2}, -1)}(z, {\overline z}))$
and $[ {\overline Q}, V_{F}^{(-{1\over 2}, -1)}(z, {\overline z})]=
{\overline{\partial}}({\overline c}({\overline z})
V_{F}^{(-{1\over 2}, -1)}(z, {\overline z}))$
when the wavefunctions $u$ and $\upsilon$ satisfy:
$$
\eqalign{
&({\gamma^{\lambda}})_{\beta}^{{\dot{\beta}}}
{\partial_{\lambda}}{\upsilon}^{{\dot{\alpha}}}_{\mu
\nu\rho}(X){\delta_{{\dot{\alpha}}{\dot{\beta}}}}
+{{\sqrt 2}\over {(D-2)}}u^{\alpha}_{\mu\nu\rho}(X)
{\delta_{{\alpha}{\beta}}}=0, \quad
{\partial^{\mu}}u^{\alpha}_{\mu\nu\rho}(X)=0, \cr
&({\gamma^{\lambda}})_{\alpha{\dot{\beta}}}
{\partial_{\lambda}}u^{\alpha}_{\mu
\nu\rho}(X)
+{(D-2)}{\upsilon}^{{\dot{\alpha}}}_{\mu\nu\rho}(X)
{\delta_{{\dot{\alpha}}{\dot{\beta}}}}=0, \quad
({\gamma^{\mu}})_{\dot{\alpha}}^{\beta}{\upsilon}^{{\dot{\alpha}}}_{\mu
\nu\rho}(X)=0,\cr
&({\gamma^{\mu}})_{{\alpha}{\dot{\beta}}}u^{\alpha}_{\mu\nu\rho}(X)
-(D-2){\partial^{\mu}}{\upsilon}^{{\dot{\alpha}}}_{\mu
\nu\rho}(X){\delta_{{\dot{\alpha}}{\dot{\beta}}}}=0,
\quad {\partial^{\nu}}u^{\alpha}_{\mu\nu\rho}(X)=
{\partial^{\nu}}{\upsilon}^{{\dot{\alpha}}}_{\mu\nu\rho}(X)=0, \cr}
\numbereq\name{\eqdyo}
$$
where $D=10$ is the space-time dimension.
Conformal invariance thus provides equations of motion and
gauge conditions for the wavefunctions $u^{\alpha}_{\mu\nu\rho}(X)$
and ${\upsilon}^{{\dot{\alpha}}}_{\mu\nu\rho}(X)$.
These constraints can be written in a compact form if we combine
$u$ and $\upsilon$ into a Dirac spinor $\psi={u \choose {(D-2){\upsilon}}}:$
$$
\eqalign{
&{\gamma^{\lambda}}{\partial_{\lambda}}{\psi}_{\mu\nu\rho}(X)+
{\sqrt 2}{\psi}_{\mu\nu\rho}(X)=0,\cr
&{\gamma^{\mu}}{\psi}_{\mu\nu\rho}(X)=
{\partial^{\mu}}{\psi}_{\mu\nu\rho}(X)=
{\partial^{\nu}}{\psi}_{\mu\nu\rho}(X)=
{\partial^{\rho}}{\psi}_{\mu\nu\rho}(X)=0.
}
\numbereq\name\eqasxcm
$$
We observe that ${\psi}_{\mu\nu\rho}(X)$ obeys a massive Dirac
equation and a set of gauge conditions.
In order to extract the couplings of massive
spin-7/2 particles to gravity we need to calculate
the three point scattering amplitude of two spin-7/2
fermions and a graviton. To satisfy the superconformal ghost
charge condition we consider the graviton vertex operator
in the $(q, {\overline q}) = (-1, 0)$ picture
$$
V_{G}^{(-1, 0)}(z, {\overline z})=h_{\mu\nu}(X){\psi}^{\mu}
e^{-{\phi}}{\overline{\partial}}X^{\nu}+{\partial_{\lambda}}
h_{\mu\nu}(X){\psi}^{\mu}
e^{-{\phi}}{\overline{\psi}}^{\lambda}{\overline{\psi}}^{\nu},
\numbereq\name{\eqardiut}
$$
where the graviton wavefunction obeys $\Box h^{\mu\nu}(X)=
{\partial_{\mu}}h^{\mu\nu}(X)=0$. We proceed now to calculate
$\langle V_{F}^{(-{1\over 2}, -1)}(z_1,
{\overline z}_1)V_{G}^{(-1, 0)}(z_2, {\overline z}_2)
V_{F}^{(-{1\over 2}, -1)}(z_3, {\overline z}_3)\rangle$.

To calculate superstring three point functions involving
spin fields we use the techniques developed in
\ref\stur{V. A. Kostelecky, O. Lechtenfeld, S. Samuel
and S. Watamura, \np288  173 (1987); S. Samuel, \np294 (1987) 365.}.
In particular, we note that three point functions always factorize
into a product of a holomorphic and an antiholomorphic piece.  We recall
that the closed string graviton vertex operator, (\eqardiut), is simply a
product of separate left- and right-moving gauge boson vertices.  Starting
with the holomorphic Ramond sector, we now calculate the three point
function of two massive fermions with a gauge boson. Dynamical issues
of massive string states in open string theory were discussed in
\ref\lab{J. M. F. Labastida and M. Vozmediano, \np312 (1989) 308;
P. C. Argyres and C. R. Nappi, \pl224 (1989) 89;
I. L. Buchbinder, V. A. Krykhtin and V. D. Pershin, \pl348 (1995) 63.}.

The vertex operators which describe massive spin-3/2 fermions
and massless gauge bosons in the $q=-{1\over 2}$ and $q=-1$
ghost pictures respectively are given by
$$
V_{F}^{(-{1\over 2})}(z, {\overline z})=u^{\alpha}_{\mu}(X)
S_{\alpha}e^{-{{\phi}\over 2}}{\partial}X^{\mu}
+{\upsilon}^{{\dot{\alpha}}}_{\mu
}(X)({\gamma}_{\lambda})_{{\dot{\alpha}}}^{\beta}{\psi}^{\mu}
{\psi}^{\lambda}S_{\beta}e^{-{{\phi}\over 2}}, \quad
V_{B}^{(-1)}(z, {\overline z})=A_{\mu}(X){\psi}^{\mu}
e^{-{\phi}}.
\numbereq\name{\eqbvna}
$$
Conformal invariance implies that $A_{\mu}$ and
$\psi={u \choose {(D-2){\upsilon}}}$ satisfy
$$
{\gamma^{\lambda}}{\partial_{\lambda}}{\psi}_{\mu}(X)+
{\sqrt 2}{\psi}_{\mu}(X)=0, \quad {\gamma^{\mu}}{\psi}_{\mu}(X)=
{\partial^{\mu}}{\psi}_{\mu}(X)=0, \quad
\Box A^{\mu}(X)=
{\partial_{\mu}}A^{\mu}(X)=0.
\numbereq\name\eqnvxhg
$$
The gauge conditions $\gamma\cdot\psi=\partial\cdot\psi=0$ eliminate
the spin-1/2 components and thus the vector-spinor
wavefunction ${\psi}^{\alpha}_{\mu}$ describes a pure spin-3/2
massive open string state.

Taking into account momentum conservation and the conditions that
different polarizations obey, we find
$$\eqalign{
{\cal A}^\gamma_{FF}(\psi_{1\mu},k_1;A_{2\sigma},k_2;\psi_{3\mu},k_3) =
{1\over {\sqrt 2}}&\left [{\overline {\psi}}_{1\mu}
{\gamma^{\sigma}}A_{2\sigma}
{\psi}_{3}^{\mu}-{\overline {\psi}}_{1\mu}
{\gamma^{\sigma}}A_{2\sigma}{\gamma_{+}}{\psi}_{3\nu}
k_2^\mu k_2^\nu \right ]\cr
+ i &\left [
{\overline {\psi}}_{1\mu}{\gamma_{+}}{\psi}_{3}^{\nu}
-{\overline {\psi}}_{1\mu}{\gamma_{-}}{\psi}_{3}^{\nu} \right ]
k_2^\mu A_2^\nu,}
\numbereq\name{\eqfgdp}
$$
where $\gamma_{+}(\gamma_{-})$ projects the Dirac spinor $\psi$
onto its positive (negative) chirality components (in the ten-dimensional
sense).

While this is meant to be viewed as the holomorphic component of the closed
string amplitude, it also has the interpretation as an open string
calculation of a single photon emission from a spin-3/2 particle.
To this linear order in the gauge field, the corresponding effective
field theoretic Lagrangian reproducing the
three point function (\eqfgdp) is given by
$$
{\cal L}_{(3/2)}=-{i\over 2}{\overline {\psi}}_{\mu}
(\gamma^\lambda D_\lambda + m){\psi}^{\mu}-{1\over {2m}}
{\overline {\psi}}_{\mu}F^{\mu\nu}{\psi}_{\nu}+
{1\over {m^2}}{\overline {\psi}}_{\mu}{\partial^{\mu}}F_{\nu\lambda}
{\gamma^{\lambda}}{\gamma_{+}}{\psi^{\nu}}+\cdots,
\numbereq\name{\eqasxcu}
$$
where $m=\sqrt2$ and $D_\mu = \partial_\mu + i A_\mu$.  The $\cdots$
represent both terms that possibly vanish on-shell and terms which
are necessary to implement the physical state constraints (including
the introduction of non-physical degrees of freedom).  In the present
letter we are mainly interested only in the form of the interaction
between massive higher-spin string states and external gravitational
or electromagnetic fields and not in the details of the additional terms
necessary for the consistency of the effective Lagrangian%
\note{See {\it e.g.}~%
\ref\ena{M. Evans and I. Giannakis, \pr44 (1991) 2467.}
for a discussion about the need to introduce non-physical degrees
of freedom for higher spin string states.}.

Completing the spin-7/2 calculation, we now examine the
antiholomorphic sector, corresponding to gauge boson emission from a massive
spin-2 boson. The calculation is straightforward, especially for the NS
sector in the covariant formalism. The corresponding vertex operators
in the ghost picture $q=0$ and $q=-1$ read
$$
V_{B}^{(0)}(z, {\overline z})=A_{\mu}(X)
{\overline{\partial}}X^{\mu}, \quad
V_{B}^{(-1)}(z, {\overline z})={\phi}_{\mu\nu}(X)e^{-{\overline{\phi}}}
{\overline{\psi}}^{\mu}
{\overline{\partial}}X^{\nu},
\numbereq\name{\eqbvmxa}
$$
while the resulting amplitude is given by
$$
\eqalign{
{\cal A}_{BB}^\gamma(\phi_{1\mu\nu},k_1;A_{2\sigma},k_2;\phi_{3\mu\nu},k_3)
= \phi_{1\mu\nu}\phi_{3\lambda\sigma}\bigl[
& \eta^{\mu\lambda}\eta^{\nu\sigma}A_2\cdot k_3
+ 4\eta^{\vphantom{[]}\mu\lambda} k_2^{[\nu}A_2^{\sigma]}\cr
&-\eta^{\mu\lambda}k_2^\nu k_2^{\sigma}A_2\cdot
k_3-2k_2^\mu k_2^\lambda k_2^{[\nu}A_2^{\sigma]}\bigr],
}
\numbereq\name{\eqspintwo}
$$
with corresponding (open string) effective Lagrangian
$$
\eqalign{
{\cal L}_{(2)}=-\half D_\rho\phi_{\mu\nu}D^\rho\phi^{\mu\nu}
-\half m^2\phi_{\mu\nu}\phi^{\mu\nu}
&+2i\phi_{\mu\nu}F^{\nu\sigma}\phi^\mu{}_\sigma\cr
&-{2i\over m^2}(\phi_{\mu\nu}\partial^\nu F_{\rho\sigma}\partial^\rho
\phi^\mu{}_\sigma-\phi_{\mu\nu}\partial^\mu\partial^\lambda F_{\nu\sigma}
\phi^\sigma{}_\lambda).}
\numbereq\name{\eqbasde}
$$

The three point function for the closed string states, ${\cal A}^h_{FF}$,
follows by combining the holomorphic and antiholomorphic three point functions
for the open string states, Eqns.~(\eqfgdp), and (\eqspintwo).  We find
$$
\kern-10pt	
\eqalign{
{\cal A}_{FF}^h(&\psi_{1\mu\nu\lambda},k_1;h_{\mu\nu},k_2;
\psi_{3\mu\nu\lambda},k_3) = {\cal A}_{FF}^{\gamma(\rm left)}
\times {\cal A}_{BB}^{\gamma(\rm right)}\cr
=&{\overline {\psi}}_{1\mu\nu\lambda}\gamma^\sigma
\psi_3^{\mu\nu\lambda}h_{\sigma\rho} k^\rho_3
+2{\overline{\psi}}_{1\mu\nu\lambda}
{\gamma^{\sigma}}{\psi}_{3}^{\mu\nu\rho}
(h_{\rho\sigma}k^\lambda_2
-h^\lambda{}_\sigma k_{2\rho}^{\vphantom{\lambda}})
+i\sqrt{2}({\overline{\psi}}_{1\mu\nu\lambda}
\gamma_+\psi_{3\sigma}^{\vphantom{\nu}}{}^{\nu\lambda}\cr
&-{\overline{\psi}}_{1\sigma\nu\lambda}
\gamma_-\psi_{3\mu}^{\vphantom{\nu}}{}^{\nu\lambda})
h^{\mu\rho} k^\sigma_2 k_{3\rho}^{\vphantom{\sigma}}
+2i\sqrt{2}({\overline{\psi}}_{1\mu\nu\lambda}\gamma_+
\psi_{3\sigma}^{\vphantom{\nu}}{}^{\nu\rho}
-{\overline{\psi}}_{1\sigma\nu\lambda}
\gamma_-\psi_{3\mu}^{\vphantom{\nu}}{}^{\nu\rho})(h^\mu{}_\rho k_2^\sigma
k_2^\lambda-h^{\mu\lambda}k_2^\sigma k_{2\rho}^{\vphantom{\nu}})\cr
&-{\overline {\psi}}_{1\mu\nu\lambda}\gamma^\sigma(1+\gamma_+)
\psi_{3\rho}^{\vphantom{\nu}}{}^{\nu\lambda}
h_{\sigma\delta}k_3^{\delta\vphantom{\mu}} k_2^\mu k_2^\rho
- {\overline {\psi}}_{1\mu\nu\lambda}\gamma^\sigma(1+2\gamma_+)
\psi_3^{\mu\rho\delta}
(h_{\sigma\delta} k_2^\nu k_{2\rho}^{\vphantom{\nu}}k_2^\lambda
- h_{\sigma}{}^{\lambda}k_2^\nu k_{2\rho}^{\vphantom{\nu}}
  k_{2\delta}^{\vphantom{\nu}})\cr
&-i\sqrt{2}({\overline{\psi}}_{1\mu\nu\lambda}
\gamma_+\psi_{3\sigma}^{\vphantom{\nu}}{}^{\nu\rho}
-{\overline{\psi}}_{1\sigma\nu\lambda}
\gamma_-\psi_{3\mu}^{\vphantom{\nu}}{}^{\nu\rho})h^{\mu\delta}
k^\sigma_2 k_{3\delta}^{\vphantom{\nu}} k_2^\lambda
k_{2\rho}^{\vphantom{\nu}}\cr
&-i\sqrt{2}({\overline{\psi}}_{1\mu\nu\lambda}
\gamma_+\psi_{3\sigma\rho\delta}^{\vphantom{\nu}}
-{\overline{\psi}}_{1\sigma\nu\lambda}
\gamma_-\psi_{3\mu\rho\delta}^{\vphantom{\nu}})
(h^{\mu\delta}  k^{\sigma\vphantom{\rho}}_2 k^{\nu\vphantom{\rho}}_2
k_2^{\lambda\vphantom{\rho}} k_2^\rho
-h^{\mu\lambda} k^{\sigma\vphantom{\rho}}_2 k^{\nu\vphantom{\rho}}_2
k_2^{\delta\vphantom{\rho}}  k_2^\rho)\cr
&+{\overline{\psi}}_{1\mu\nu\lambda}\gamma^\sigma
\gamma_+\psi_{3\rho}^{\vphantom{\nu}}{}^{\nu\delta}h_{\sigma\epsilon}
k^{\epsilon\vphantom{\mu}}_3 k^\mu_2 k^{\lambda\vphantom{\mu}}_2
k_{2\delta}^{\vphantom{\mu}} k_2^\rho
+{\overline{\psi}}_{1\mu\nu\lambda}\gamma^\sigma
\gamma_+\psi_{3\rho\delta\epsilon}^{\vphantom{\nu}}
(h_{\sigma}{}^{\epsilon} k^{\nu\vphantom{\mu}}_2 k^\mu_2
k^{\lambda\vphantom{\mu}}_2  k_2^{\delta\vphantom{\mu}} k_2^\rho
-h_{\sigma}{}^{\lambda}  k^{\nu\vphantom{\mu}}_2 k^\mu_2
k^{\epsilon\vphantom{\mu}}_2 k_2^{\delta\vphantom{\mu}} k_2^\rho)
.\cr}
\kern-23pt	
\numbereq\name{\eqanvc}
$$
As a result the effective Lagrangian that reproduces this particular
three point function contains terms up to five derivatives.
Up to terms that vanish on-shell (but that are nevertheless
crucial for the consistency of such a massive higher-spin Lagrangian), we
find
$$
\eqalign{
{\cal L}_{\rm eff}={\sqrt {-g}}&[{1\over 2}{\overline{\psi}}_{\mu\nu\lambda}
({\gamma^{\rho}}{\nabla_{\rho}}+m){\psi}^{\mu\nu\lambda}
-{1\over m} {\overline{\psi}}_{\mu\nu\lambda}(R^{\mu\alpha\nu\beta}
-{1\over 8}
R^{\mu\alpha\rho\sigma}{\gamma_{\rho\sigma}}{\eta}^{\nu\beta})
{\psi}_{\alpha\beta}{}^{\lambda}\cr
&-{1\over {m^2}}(
{\overline{\psi}}_{\mu\nu\lambda}R^{\sigma(\mu\alpha)\rho}{\gamma}_{
\sigma}(1+{\gamma_{+}}){\partial_{\rho}}{\psi}_{\alpha}{}^{\nu\lambda}
-{\overline{\psi}}_{\mu\nu\lambda}{\partial^{(\mu}}
R^{\alpha)\sigma\nu\beta}{\gamma_{\sigma}}(1+2{\gamma_{+}})
{\psi}_{\alpha\beta}{}^{\lambda})\cr
&-{1\over {m^3}}({\overline{\psi}}_{\mu\nu\lambda}{\partial^{(\nu}}
R^{\beta)\sigma\mu\alpha}{\partial_{\sigma}}
{\psi}_{\alpha\beta}{}^{\lambda}+{\overline{\psi}}_{\mu\nu\lambda}
{\partial^{\mu}}{\partial^{\alpha}}R^{\nu\beta\lambda\gamma}
{\psi_{\alpha\beta\gamma}})\cr
&-{2\over {m^4}}({\overline{\psi}}_{\mu\nu\lambda}{\partial^{\mu}}
{\partial^{\alpha}}R^{\sigma(\beta\nu)\delta}{\gamma_{\sigma}}
{\gamma_{+}}{\partial_{\delta}}{\psi_{\alpha\beta}{}^{\lambda}}
-{\overline{\psi}}_{\mu\nu\lambda}{\partial^{\nu}}{\partial^{\delta}}
{\partial^{(\lambda}}R^{\gamma)\sigma\mu\rho}{\gamma_{\sigma}}
{\gamma_{+}}{\psi_{\rho\delta\gamma}})].\cr}
\numbereq\name{\eqdanhi}
$$
Several points are in order here.  First of all, working with the three
point function, we only obtain information up to linearized order in the
graviton, $h_{\mu\nu}$.  For this reason, at this order there is no
distinction between bare and covariant derivatives of the Riemann tensor
that appear in (\eqdanhi).  Secondly, this effective Lagrangian is by no
means unique, as we are always allowed to shift it by terms vanishing
on-shell.  In particular, we note that the $\gamma$-transverse condition
$\gamma\cdot\psi=0$ allows use of the on-shell identity
$$
\overline{\psi}_{\mu\nu\lambda}R^{\mu\alpha\rho\sigma}\gamma_{\rho\sigma}
\eta^{\nu\beta} \psi_{\alpha\beta}{}^\lambda=2\overline{\psi}_{\mu\nu\lambda}
R^{\mu\alpha\nu\beta}\psi_{\alpha\beta}{}^\lambda
-\overline{\psi}_{\mu\nu\lambda}R^{\mu\alpha\rho\sigma}
\gamma_{\rho\sigma}{}^{\nu\beta}\psi_{\alpha\beta}{}^\lambda,
\numbereq\name{\eqpapadia}
$$
indicating that even the leading two-derivative term is by no means unique.
Furthermore, note that the presence of the ten-dimensional chirality
projection $\gamma_+$ for spacetime fermions indicates that, even while
massive, such states maintain chiral interactions with gravity.  It is at
this point where the difference between massive IIA and IIB string states
shows up.  The complete Type II spectrum includes in fact a pair of spin-7/2
states at the first mass level, one each from the R-NS and the NS-R sectors.
In our conventions, Eqn.~(\eqdanhi) corresponds to the state from the R-NS
sector, while a similar equation with either $\gamma_-$ or $\gamma_+$ would
describe the state from the NS-R sector for the Type IIA or IIB theory
respectively.

Finally, the effective gravitational interactions of
string states at higher mass levels have correspondingly higher derivative
couplings.  Physically, this corresponds to the intuitive notion that
highly excited strings are spread out, and hence feel tidal effects
arising from the curvature of spacetime. This departure from the
minimal coupling prescription leads to violation of the
strong equivalence principle
\ref{\gast}{F. A. Berends and R. Gastmans, \ap98 (1976) 225;
K. A. Milton, \pr15 (1977) 2149;
I. T. Drummonds and S. J. Hathrell, \pr22 (1980) 343;
R. D. Daniels and G. M. Shore, \np425 (1994) 634;
P. F. Mende, in {\it String Quantum Gravity and
Physics at the Planck Scale}, Proceedings of the Erice Workshop,
1992, edited by N. Sanchez (World Scientific, Singapore, 1993);
R. Lafrance and R. C. Myers, \pr51 (1995) 2584.}.
This is just a fact of life; in relativistic quantum field theory particles
have an intrinsic size: their Compton wavelength. This makes them behave
in some respect as extended objects, sensitive to tidal forces.
\newsection Non-minimal coupling and the gravitational quadrupole.%
Non-minimal couplings to the Riemann tensor were discussed in Ref.~[\porrati]
in the context of point particle field theory, where it was shown that
tree-level unitarity for particles of spin $>2$ demands the presence of just
such a non-minimal term.  The authors of [\porrati] give a general
expression for the required non-minimal addition to the action for both
integer and half-integer spins.  For the former, the on-shell Lagrangian
(also ignoring auxiliary fields)%
\note{See {\it e.g.}~%
\ref\hagen{L. P. S. Singh and C. R. Hagen, \pr9 (1974) 898, 910.}
for an explicit form of the massive higher spin Lagrangian.}
for a boson of spin $s$ takes the form
$$
{\cal L}=\half\phi^{(s)}(\nabla_\mu\nabla^\mu-m^2)\phi^{(s)}
+{s(s-1)\over2}\phi^{(s)}_{\mu\nu\ldots}R^{\mu\alpha\nu\beta}
\phi^{(s)}_{\alpha\beta}{}^{\ldots}+\cdots.
\numbereq\name{\eq0anou}
$$
The resulting equation of motion for $\phi^{(s)}$ may then be expressed as
$$
(\nabla_\mu\nabla^\mu-m^2)\phi^{(s)}
+[R_{\mu\nu\lambda\sigma}\half\Sigma^{\mu\nu}\half\Sigma^{\lambda\sigma}]
\phi^{(s)}+\cdots=0,
\numbereq \name\eqbquad
$$
where $\cdots$ indicate terms vanishing on-shell, and $\Sigma^{\mu\nu}$ are
the Lorentz generators in the spin-$s$ representation,
$$
(\Sigma^{\mu\nu})_{\{\alpha\}}{}^{\{\beta\}}=
2s\delta^{[\mu}{}_{[\alpha_1}^{\vphantom{[]}}\eta^{\nu][\beta_1}
\delta^{\beta_2\ldots\beta_s]}_{\alpha_2\ldots\alpha_s]},
\numbereq\name{\eqfion}
$$
where all symbols are antisymmetric with weight one.

On the other hand, for spin-$(n+\ha)$ fermions in four dimensions, the
non-minimal Lagrangian is
$$
{\cal L}=\overline{\psi}^{(n)}(\gamma^\mu\nabla_\mu-m)\psi^{(n)}
+{n(n-1)\over2m}\overline{\psi}^{(n)}_{\mu\nu\ldots}
R^{+\,\mu\alpha\nu\beta}\psi^{(n)}_{\alpha\beta}{}^{\ldots}+\cdots
\numbereq\name\eqfirstord
$$
(again only up to terms vanishing on-shell).  Here
$R_{\mu\nu\lambda\sigma}^+ = R_{\mu\nu\lambda\sigma}+\ha R_{\mu\nu\alpha\beta}
\gamma^{\alpha\beta}{}_{\lambda\sigma}$ is a feature of the four-dimensional
theory.  Taking the first order equation of motion from (\eqfirstord)
and multiplying by $(\gamma^\mu\nabla_\mu+m)$, we obtain the second order
equation
$$
(\nabla_\mu\nabla^\mu-m^2)\psi^{(n)}
+[R_{\mu\nu\lambda\sigma}\half\Sigma^{\mu\nu}\half\Sigma^{\lambda\sigma}]
\psi^{(n)}+\cdots=0,
\numbereq\name\eqfquad
$$
where this time
$$
(\Sigma^{\mu\nu})_{\{\alpha\}}{}^{\{\beta\}}=\half\gamma^{\mu\nu}
\delta^{[\beta_1\ldots\beta_n]}_{[\alpha_1\ldots\alpha_n]}
+2n\delta^{[\mu}{}_{[\alpha_1}^{\vphantom{[]}}\eta^{\nu][\beta_1}
\delta^{\beta_2\ldots\beta_n]}_{\alpha_2\ldots\alpha_n]}
\numbereq\name{\eqkokot}
$$
is the Lorentz generator in the spin-$(n+\ha)$ representation.

We now see from (\eqbquad) and (\eqfquad) that both integer and
half-integer spin fields have identical forms for the preferred
non-minimal coupling resulting from tree-level unitarity concerns.
Furthermore, since the Riemann coupling has the form of a gravitational
quadrupole moment, these results of Ref.~[\porrati] are suggestive of a
gravitational version of the corresponding statement of ``g=2'' as a
natural value for the gyromagnetic ratio for electromagnetic couplings
\ref{\por}{S. Ferrara, M. Porrati and V. Telegdi, \pr46 (1992) 3529.}.

The above discussion suggests the definition of a gravitational quadrupole
``$h$-factor'' that may be determined from the equations of motion
according to
$$
(\nabla_\mu\nabla^\mu-m^2)\varphi
+h[R_{\mu\nu\lambda\sigma}\half\Sigma^{\mu\nu}\half\Sigma^{\lambda\sigma}]
\varphi+\cdots=0.
\numbereq\name{\eqawer}
$$
While the $h$-factor may equally well be defined in terms of the
non-minimal coupling of $\varphi$ to the Riemann tensor in the
Lagrangian, such a definition is complicated by the fact that there is an
inherent ambiguity in the minimal coupling prescription itself (which is not
present for the equations of motion).  Using this definition, the results of
Ref.~[\porrati] may be concisely summarized by the statement that $h=1$ is
the preferred value of the $h$-factor based on the above field theory
considerations.

Turning to the spin-7/2 Lagrangian, (\eqdanhi), it is clear that it
cannot be written in the ``preferred'' form of (\eqfirstord), even through
the use of the on-shell manipulation (\eqpapadia)%
\note{The factor of $1\over2$ arises in (\eqdanhi) because there
$\psi_{\mu\nu\lambda}$ is a Majorana spinor, while in (\eqfirstord)
$\psi^{(n)}$ is a Dirac spinor.}.
As a result this provides evidence that $h\ne1$ for massive string states in
general.  In order to determine the spin-7/2 $h$-factor, we first make
use of (\eqpapadia) and work in four dimensions to note that the second
order equation of motion arising from (\eqdanhi) has the form
$$
(\nabla_\rho\nabla^\rho-m^2)\psi^{\mu\nu\lambda}
+3(R^{\mu\alpha\nu\beta}+\half R^{\mu\alpha\rho\sigma}\gamma_{\rho\sigma}
\eta^{\nu\beta})\psi_{\alpha\beta}{}^\lambda+\cdots=0.
\numbereq\name{\eqhfone}
$$
On the other hand, for spin 7/2, (\eqawer) gives instead
$$
(\nabla_\rho\nabla^\rho-m^2)\psi^{\mu\nu\lambda}
+6h(R^{\mu\alpha\nu\beta}+{\textstyle{1\over4}}
R^{\mu\alpha\rho\sigma}\gamma_{\rho\sigma}
\eta^{\nu\beta})\psi_{\alpha\beta}{}^\lambda+\cdots=0,
\numbereq\name{\eqhftwo}
$$
which clearly has a different Lorentz structure.  What this indicates is
that, even when restricted to on-shell interactions, there are in fact
{\it two} possible distinct Lorentz-invariant and parity conserving
interactions that may be written in terms of the Riemann coupling.  Thus a
single ``$h$-factor'' is insufficient, and in fact two parameters are
necessary to fully characterize this lowest order non-minimal interaction.

On the other hand, working in the Newtonian limit, we find that both
$R^{\mu\alpha\nu\beta}\psi_{\alpha\beta\ldots}$ and
$(R^{\mu\alpha\rho\sigma}\gamma_{\rho\sigma}\eta^{\nu\beta})
\psi_{\alpha\beta\ldots}$ reduce to the same form, related to the
(non-relativistic) quadrupole moment $Q^{ij}$.  In particular, in
four dimensions, the components of the Riemann tensor are given in terms of
the Newtonian potential $\phi$ as
$$
\eqalign{
R_{0i0j}&=\partial_i\partial_j\phi,\cr
R_{ijkl}&=
\delta_{ik}\partial_j\partial_l\phi+\delta_{jl}\partial_i\partial_k\phi
-\delta_{il}\partial_j\partial_k\phi-\delta_{jk}\partial_i\partial_l\phi.
}
\numbereq
$$
Furthermore, non-relativistically, the transverse and $\gamma$-transverse
conditions on $\psi_{\mu\nu\ldots}$ give both $\psi_{0\ldots} =
O(p/M)\psi_{i\ldots}\ll1$ and $\gamma^i\psi_{i\ldots}\ll1$.  Thus we find
$$
R^{\mu\alpha\nu\beta}\psi_{\alpha\beta\ldots} \to
R_{ikjl}\psi^{kl\ldots}=-2\partial_{(i}\partial^k\phi\, \psi_{j)}{}^{k\ldots}
+\delta_{ij}\partial_k\partial_l\phi\,\psi^{kl\ldots},
\numbereq
$$
where the last term, having a trace form, does not contribute diagonally to
the leading spin-7/2 quadrupole interaction, but instead gives an
off-diagonal interaction between spins 7/2 and 3/2.  For the other
possibility, we find instead
$$
(R^{\mu\alpha\rho\sigma}\gamma_{\rho\sigma}\eta^{\nu\beta})
\psi_{\alpha\beta\ldots}\to
R_{iklm}\gamma^{lm}\psi^{k\ldots}=-4\partial_i\partial_k\phi\,\psi^{k\ldots}
+2\gamma^i(\partial_k\partial_l\phi\,\gamma^k\psi_{l\ldots}).
\numbereq
$$
This time the last term (having a $\gamma$-trace form) corresponds
to an off-diagonal interaction between spins 7/2 and 5/2.

With the above in mind, in practice we define the ``$h$-factor'' of
(\eqawer) only {\it in the Newtonian limit} (in other words focusing on
angular momentum and not Lorentz generators).  As a result, the diagonal
spin-7/2 equation of motion, (\eqhfone), reduces to
$$
(\nabla_\rho\nabla^\rho-m^2)\psi^{ijk}
-6\partial^{(i}\partial_l\phi\,\psi_l{}^{jk)}+\cdots=0,
\numbereq
$$
where $\cdots$ now also include off-diagonal interactions with lower-spin
states (that are always present and fall on the sub-leading Regge
trajectories).  Contrasting this with
$$
(\nabla_\rho\nabla^\rho-m^2)\psi^{ijk}
-9h\partial^{(i}\partial_l\phi\,\psi_l{}^{jk)}+\cdots=0,
\numbereq
$$
which follows from (\eqhftwo), finally allows us to determine that $h=2/3$
for this particular massive spin-7/2 string state.

At this point we must clarify an apparent paradox. The non-relativistic
formula for the quadrupole moment is, in space-time dimension $D$:
$$
Q^{ij}=\int d^{D-1}x\left[ (D-1)x^ix^j -\delta^{ij}x^2\right]T^{00},
\qquad i,j =1,..,D-1.
\numbereq\name{\quadr}
$$
The action of particles of spin $s>1/2$ can have a non-minimal
coupling proportional to the Einstein tensor:
$$
\Delta S =\int d^Dx B^{\mu\nu}G_{\mu\nu},
\numbereq\name{\ricci}
$$
where $B^{\mu\nu}$ is a bilinear in the spin-$s$ field.
To linear order in the gravitational field, this induces the following change
in the stress energy tensor:
$$
T^{00}=\partial_i\partial_j B^{ij}+\cdots,\qquad
T^i_i=(3-D)\partial_i\partial_j B^{ij}+\cdots.
\numbereq\name{\change}
$$
Here $\cdots$ stands for terms that do not contribute to the gravitational
quadrupole.
Substituting this equation into formula~(\quadr) we obtain a non-zero change
of the quadrupole, induced by terms that vanish on the Einstein
shell!%
\note{Terms proportional to the scalar curvature tensor do not
contribute to the quadrupole.}
The solution to this paradox is that equation~(\quadr) is a good definition
of the quadrupole only for non-relativistic matter. In more general cases
this definition is wrong, {\it even in the Newtonian limit}. The correct
definition is obtained by computing the energy of the
particle in a static, slowly-varying external Newtonian potential $\phi$.
The change in energy due to the quadrupole is:
$$
\Delta H={1\over 2(D-1)}Q^{ij}\partial_i\partial_j\phi.
\numbereq\name{\energ}
$$
This formula is insensitive by construction to all terms that vanish on the
Einstein shell.
In the Newtonian limit, the metric is $g_{00}=-1 - 2\phi$,
$g_{ij}=(1 -{2\over D-3}\phi)\delta_{ij}$, all other terms vanish.
This induces the following change in the Energy:
$$
\Delta H=\int d^{D-1}x \phi
\left(T^{00} + {1\over D-3}T^{ii}\right).
\numbereq\name{\newt}
$$
Thus, the correct expression for the quadrupole is obtained by replacing
$T^{00}$ with $T^{00}+(D-3)^{-1}T^i_i$ in Eqn.~(\quadr). By substituting
Eqn.~(\change) into this new expression we find that the contribution to
the quadrupole due to non-minimal coupling to the Ricci tensor vanishes, as
it should. Eqn.~(\quadr) can be used in the Newtonian limit
{\it only} when $T^i_i \approx 0$, as it holds, for instance, in
macroscopic non-relativistic matter.

\newsection Generalization to higher-spin.%
The results for the massive spin-7/2 state are easily generalized to
arbitrary massive higher spin string states interacting with a graviton.  As
usual, the three point function factorizes into separate holomorphic and
antiholomorphic parts.  For simplicity we focus on states on the leading
Regge trajectory, namely states of spin $n+\ha$ and $n+1$ respectively for
the R and NS sectors at mass level $n$.

In the Ramond sector, the spin-$(n+\ha)$ vertex operator is given by
$$
\eqalign{
V_{F}^{(-{1\over 2})}(z, {\overline z})=u^{\alpha}_{\mu_1 \ldots \mu_n}
(X)S_{\alpha}e^{-{{\phi}\over 2}}&{\partial}X^{\mu_1} \cdots
{\partial}X^{\mu_n}\cr
&+{\upsilon}^{{\dot{\alpha}}}_{\mu_1
\ldots \mu_n}(X)({\gamma}_{\lambda})_{{\dot{\alpha}}}^{\beta}
{\psi}^{\mu_1}
{\psi}^{\lambda}S_{\beta}e^{-{{\phi}\over 2}}\partial X^{\mu_2}
\cdots {\partial} X^{\mu_n}.\cr}
\numbereq\name{\eqamb}
$$
The resulting three point function, ${\cal A}_{FF}^\gamma$, contains terms
up to $O((\alpha'k^2)^n)$.  To lowest order, we find
$$
\eqalign{
{\cal A}_{FF}^\gamma(\psi_1,k_1;A_{2\sigma},&k_2;\psi_3,k_3)
= {n!\over\sqrt{2}}\bigl[
\overline\psi_{1\mu_1\ldots\mu_n}A_{2\sigma}
\gamma^{\sigma}\psi_3^{\mu_1\ldots\mu_n}\cr
& -i{2n\over m}(\overline\psi_{1\alpha\mu_2\ldots\mu_n}\gamma_-
\psi_{3\beta}{}^{\mu_2\ldots\mu_n}-
\overline\psi_{1\beta\mu_2\ldots\mu_n}\gamma_+
\psi_{3\alpha}{}^{\mu_2\ldots\mu_n})k_2^\alpha k_2^\beta+\cdots\bigr],
}
\numbereq\name{\eqargyr}
$$
where $m=\sqrt{2n}$ is the mass of the $n^{\rm th}$ excited level.

Prior to examining the closed string graviton amplitude, we note that the
effective field theoretic Lagrangian which reproduces this three point
function for the open string may be written as
$$
{\cal L}_{n+\ha}^{(\rm open)}
=-{i\over 2}{\overline {\psi}}_{\mu_{1} \cdots \mu_{n}}
(\gamma^\sigma D_\sigma+m)
{\psi}^{\mu_{1} \cdots \mu_{n}}-{1\over {2m}}{\sum_{i=1}^{n}}
{\overline {\psi}}_{\mu_{1} \cdots \mu_{n}}F^{\mu_{i}\nu_{i}}
\psi^{\mu_1\ldots\mu_i\kern-7pt/\kern3pt\ldots\mu_n}_{\hphantom{\mu_1\ldots}
\nu_i}
+ \cdots.
\numbereq \name{\eqxcvxu}
$$
{}From the form of the Lagrangian we can verify that the
the gyromagnetic ratio $g$ for all such massive fermionic open
string states on the leading Regge trajectory is equal to $2$ [\por].

For the spin-$(n+1)$ state in the NS sector, we find on the other hand the
leading behavior for the three point function
$$
\eqalign{
{\cal A}_{BB}^\gamma(\xi_1,k_1;\zeta_2,k_2;\xi_3,k_3)
= n!\xi_{1\alpha\mu_1\mu_2\ldots\mu_n}\xi_{3\beta\nu_1}{}^{\mu_2\ldots\mu_n}
&\zeta_{2\lambda}\bigl[
(\eta^{\alpha\beta}k_3^\lambda-\eta^{\alpha\lambda}k_2^\beta
+\eta^{\beta\lambda}k_2^\alpha)\eta^{\mu_1\nu_1}\cr
&+n(\eta^{\nu_1\lambda}k_2^{\mu_1}
-\eta^{\mu_1\lambda}k_2^{\nu_1})\eta^{\alpha\beta}+\cdots\bigr],}
\numbereq\name{\eqargyrhs}
$$
where once again $m=\sqrt{2n}$.  In general, as for the fermions, the
complete expression contains terms up to $O((\alpha'k^2)^n)$.
Viewed as an open string amplitude, the corresponding effective Lagrangian
has the form
$$
\eqalign{
{\cal L}_{n+1}^{(\rm open)}=
-\half D_\lambda\phi_{\mu_0\ldots\mu_n}D^\lambda\phi^{\mu_0\ldots\mu_n}
-\half m^2&\phi_{\mu_0\ldots\mu_n}\phi^{\mu_0\ldots\mu_n}\cr
&+i\sum_{i=0}^n\phi_{\mu_0\ldots\mu_n}F^{\mu_i\nu_i}
\phi^{\mu_0\ldots\mu_i\kern-7pt/\kern3pt\ldots\mu_n}_{\hphantom{\mu_0\ldots}
\nu_i}+\cdots.}
\numbereq\name{\eqvefhs}
$$
As expected, this indicates that the non-minimal electromagnetic coupling
to $F_{\mu\nu}$ gives precisely $g=2$ for the open string.

Turning now to the closed string, combining (\eqargyr) and
(\eqargyrhs) and symmetrizing
on the vector indices (corresponding to the maximal spin state at mass
level $n$) gives the following on-shell form of the effective Lagrangian:
$$
\eqalign{
{\cal L}_{2n+{3\over2}}&=\sqrt{-g}\bigl[
\half\overline\psi^{(2n+1)}(\gamma^\mu\nabla_\mu+m)\psi^{(2n+1)}\cr
&-{n(n+1)\over2m}\overline\psi_{\mu_0\mu_1\ldots\mu_{2n}}
(R^{\mu_0\nu_0\mu_1\nu_1}-{1\over4(n+1)}R^{\mu_0\nu_0\lambda\sigma}
\gamma_{\lambda\sigma}
\eta^{\mu_1\nu_1})\psi_{\nu_0\nu_1}{}^{\mu_2\ldots\mu_{2n}}
+\cdots].}
\numbereq\name{\eqpouitr}
$$
The resulting second order equation for $\psi$ takes the form
$$
\eqalign{
(\nabla_\rho\nabla^\rho-m^2)\psi_{\mu_0\mu_1\ldots}
+\bigl[&2(S_L-\half)(S_R-\half)R^{\mu_0\nu_0\mu_1\nu_1}\cr
&+\half(S-\half)R^{\mu_0\nu_0\lambda\sigma}\gamma_{\lambda\sigma}
\eta^{\mu_1\nu_1}\bigr]\psi_{\nu_0\nu_1\ldots}+\cdots=0,}
\numbereq
$$
where $S_L = n+1$ and $S_R=n+\ha$ are the components of the spin
contributed by left- and right-movers on the worldsheet respectively, and
$S=S_L+S_R$ ($=2n+{3\over2}$) is the total (spacetime) spin.  Using
this suggestive form of the non-minimal interaction, we find the
corresponding $h$-factor to be
$$
h={2S_LS_R\over(S-\ha)^2}\qquad\hbox{(half-integer spins)}.
\numbereq\name\eqquadf
$$
A similar calculation in the NS-NS sector (also on the leading Regge
trajectory) gives similarly
$$
h={2S_LS_R\over S(S-1)}\qquad\hbox{(integer spins)}.
\numbereq\name\eqquadb
$$
Note the resemblance to the string $g$-factor result%
\ref\russo{J. G. Russo and L. Susskind, \np437 (1995) 611; A.
Sen, \np440 (1995) 421.}
$$
g_L=2{\langle S_z^R\rangle\over S_z},
\qquad
g_R=2{\langle S_z^L\rangle\over S_z}.
\numbereq\name{\eqaedciov}
$$
Based on the factorization of the graviton three point function in terms of
holomorphic and antiholomorphic gauge boson amplitudes and the $g=1$ result
for {\it all} massive open string states, it is now apparent that the
$h$-factor result, (\eqquadf) and (\eqquadb), is equally valid for all
massive string states and is not restricted to those on the leading
Regge trajectory.

\newsection Conclusions.%
In this letter we have examined the three-point couplings of massive
higher-spin string states with gravity.  Focusing on the spin-7/2 state at
the first mass level of the Type II string, we have obtained an effective
Lagrangian reproducing all {\it on-shell} interactions to linearized order
in the graviton.  In particular, this effective Lagrangian contains a
non-minimal two-derivative coupling of the form $(\overline\psi R \psi)$,
which was examined by the authors of [\porrati] in the context of tree-level
unitarity.  Since this has the form of a gravitational quadrupole
interaction, we have defined the gravitational ``$h$-factor'' (in analogy
with the electromagnetic $g$-factor) and demonstrated that Ref.~[\porrati]
gives $h=1$ as a preferred value of the $h$-factor in field theory.

On the other hand, in a string theory, $h$ is determined from a
combination of left- and right-moving components of the spin for
massive closed string states.  Although generically $h\ne1$ in string
theory in contrast to the field theoretic result, this is certainly
not a disaster for string theory. In field theory the tree-level
unitarity results hold for a single massive higher spin particle
interacting with gravity, while in string theory tree-level unitarity
is achieved not only by the three-point interaction but also because
a whole tower of states of arbitrarily large masses and
spins propagate as intermediate states.

\noindent{\bf Acknowledgments.} \vskip .01in \noindent
I.G. would like to thank S. Samuel for useful discussions. M.P.
would like to thank S. Deser for illuminating notes on the correct
definition of the gravitational quadrupole, that lay too long in
a drawer.  This work was supported in part by the Department of
Energy under Contract Number DE-FG02-91ER40651-TASK B, and by NSF
under grant PHY-9722083.

\immediate\closeout1
\bigbreak\bigskip

\line{\twelvebf References. \hfil}
\nobreak\medskip\vskip\parskip

\input refs

\vfil\end

\bye